# Raman evidence for Orbiton-Mediated Multiphonon Scattering in Multiferroic TbMnO$_3$

Pradeep Kumar[1], Surajit Saha[1], D V S Muthu[1], J R Sahu[2], A K Sood[1,2,*] and C N R Rao[2]

[1]Department of Physics, Indian Institute of Science, Bangalore-560012, India

[2]Chemistry and Physics of Materials Unit and International Centre for Materials Science, Jawaharlal Nehru Centre for Advanced Scientific Research, Bangalore-560064, India

## Abstract

Temperature-dependent Raman spectra of TbMnO$_3$ from 5 K to 300 K in the spectral range of 200 to 1525 cm$^{-1}$ show five first-order Raman allowed modes and two high frequency modes. The intensity ratio of the high frequency Raman band to the corresponding first order Raman mode is nearly constant and high ( ~ 0.6 ) at all temperatures, suggesting a orbiton-phonon mixed nature of the high frequency mode. One of the first order phonon modes shows anomalous softening below T$_N$ (~ 46 K), suggesting a strong spin-phonon coupling.

# 1. Introduction

The effect of orbital ordering on the Raman spectra of perovskite manganites, $RMnO_3$ (R = Rare Earth), has been investigated both theoretically [1-4] and experimentally [3, 5-8]. Three broad bands near 1000 cm$^{-1}$, 1170 cm$^{-1}$ and 1290 cm$^{-1}$ in the Raman spectra of $LaMnO_3$ were attributed to orbiton excitations [3, 9], an assignment still being debated and alternative proposals made [1-2, 4, 10]. Although in centro-symmetric $LaMnO_3$, Raman modes are not infrared active, infrared absorption [10] shows similar bands as in the Raman spectra, attributing these features to multiphonon scattering instead of orbital excitations. As a result of strong electron-phonon coupling, Allen et al [1] have proposed that orbitons in $LaMnO_3$ are self-trapped by the local rearrangement of the lattice and hence multiphonon Raman scattering with intensities comparable to the one phonon Raman modes has been predicted. This arises from the Franck-Condon (FC) process via the self-trapped orbitons, suggesting a mixed character of phonons and orbitons to the high frequency modes. This mixed character has also been shown theoretically by other calculations considering the effects of super-exchange and electron-phonon interactions [11, 4]. The high intensity ratio (~ 0.1 to 0.4) of the second order modes to their first order counterparts has been observed experimentally for $LaMnO_3$ [5] and $RMnO_3$ (R = La, Pr, Ho and Y) [6], supporting the theoretical proposal for the mixed nature of the multiphonon bands. On the other hand, a recent room temperature Raman study of $RMnO_3$ (R = Pr, Eu, Dy, Ho and Y) and $O^{18}$ isotopically substituted $EuMnO_3$ [12] suggests that the high frequency modes are due to second order scattering involving only Brillouin zone boundary phonons. All these experimental studies have been carried out at room temperature and above. Our present Raman study looks at multi-ferroic $TbMnO_3$ as a function of temperature from 5K to 300K, covering spectral range from 200 cm$^{-1}$ to 1525 cm$^{-1}$ and focuses mainly on the temperature dependence of the two high energy excitations observed at 1168 cm$^{-1}$ and 1328 cm$^{-1}$.

$TbMnO_3$ is orthorhombic (space group Pbnm) at room temperature and shows an incommensurate lattice modulation at $T_N$ for sinusoidal antiferromagnetic ordering (with $T_N \sim$ 41K [13] or $T_N \sim$ 46K as reported by Bastjan et al. [14] ). Ferroelectric order develops at the incommensurate-commensurate transition temperature $T_{FE} \sim 27K$ [13]. As the temperature is further lowered, rare-earth $Tb^{3+}$ ion-spins also order anti-ferromagnetically at $\sim 7K$ [13]. In recent years, focus in multiferroics has been on the electro-magnons i.e magnons with an electric dipole excited by an applied ac electric field, observed below 100 cm$^{-1}$ [15-19]. As far as first-order phonons are concerned, it has been shown that in RMnO$_3$ (R = La, Nd, Sm, Gd, Dy, Pr and Tb), a few Raman and IR phonons involving oxygen vibrations are anomalous i.e. the phonon frequency decreases as temperature is lowered below $T_N$ [20-22, 23] arising from spin-phonon coupling. There have been reports of Raman studies on multi-ferroic TbMnO$_3$ dealing with only first-order Raman scattering [20, 24-26], but to our knowledge, there is no report of the high frequency excitations in TbMnO$_3$. In this paper, we present Raman scattering data from an unoriented single crystal of TbMnO$_3$ as a function of temperature in the range 5 K to 300 K. We show that the intensity ratio of the second order mode to the corresponding first order is very high and remains constant with temperature as predicted by Allen et al [1-2] for coupled multiphonons-orbiton modes. In addition, the first order mode involving oxygen vibrations ($\omega \sim 616$ cm$^{-1}$) shows anomalous softening below $T_N$ possibly due to strong spin-phonon coupling.

## 2. Experimental Details

Single crystals of TbMnO$_3$ were grown by the float-zone technique as described in ref. 23. Confocal unpolarised micro-Raman spectroscopic measurements were performed at low temperatures in backscattering geometry, using 50x long distance objective and 514.5 nm line of an Ar-ion Laser (Coherent Innova 300), covering the spectral range of 200 to 1525 cm$^{-1}$. Temperature scanning from 5 K to 300 K was done using continuous flow liquid helium cryostat with a temperature accuracy of ± 0.1 K. The scattered light was analyzed using a Raman spectrometer (DILOR XY) coupled to liquid nitrogen cooled CCD, with pixel resolution of 0.85 cm$^{-1}$ and instrumental broadening of $\sim 5$ cm$^{-1}$.

## 3. Results and Discussion

Figure 1 shows Raman spectrum at 5K displaying 8 modes labeled as S1 to S8. The spectra are fitted to a sum of Lorentzian and the frequencies, linewidths and intensities so obtained are shown in Fig. 2 for the first order Raman modes and in Fig. 3 for the multiphonon modes S7 and S8. TbMnO$_3$ is orthorhombic (space group Pbnm), with 24 Raman active modes classified as $\Gamma_{Raman} = 7A_g + 5B_{1g} + 7B_{2g} + 5B_{3g}$ [25]. The assignment of low frequency modes S1- S5, given in Table I, has been done following the work of Iliev et. al for TbMnO$_3$ [25]. The origin of mode S6 may be similar to that of the 640 cm$^{-1}$ mode observed in LaMnO$_3$ [27-28], attributed to the disorder-induced phonon density of states [12, 29] or second order Raman scattering [27]. It can also be a disorder-induced infrared active phonon mode (transverse optic mode at 641 cm$^{-1}$ and longitudinal optic mode at 657 cm$^{-1}$) observed in infrared studies of TbMnO$_3$ [23].

**(A) Temperature dependence of the first order modes**

We now discuss the temperature dependence of the modes S1 to S5 and S6. In general, the temperature dependent behavior of a phonon mode of frequency '$\omega$' is given as [21]

$$\omega(T) = \omega(0) + (\Delta\omega)_{qh}(T) + (\Delta\omega)_{anh}(T) + (\Delta\omega)_{el\text{-}ph}(T) + (\Delta\omega)_{sp\text{-}ph}(T) \qquad (1)$$

$(\Delta\omega)_{qh}(T)$ corresponds to the change in phonon frequency due to a change in the lattice parameters of the unit cell, termed as quasi-harmonic effect. $\Delta\omega_{anh}(T)$ gives the intrinsic anharmonic contributions to the phonon frequency. The effect of renormalization of the phonon frequency (($\Delta\omega)_{el\text{-}ph}(T)$) due to electron-phonon coupling is absent in insulating TbMnO$_3$. The last term, $\Delta\omega_{sp\text{-}ph}(T)$, represents the change in phonon frequency due to spin-phonon coupling, caused by the modulation of the exchange integral by lattice vibrations [21]. The change in phonon frequency of mode "$i$" due to the change in lattice parameters, i.e $(\Delta\omega)_{qh}(T)$, can be related to the change in volume if we know Grüneisen parameter $\gamma_i = -(B_0/\omega_i)(\partial\omega_i/\partial P)$, where $B_0$ is the bulk modulus and $\partial\omega_i/\partial P$ is the pressure-derivative of the phonon frequency. For a cubic crystal or isotropically expanded lattice, the change in phonon frequency due to change in volume is given as $(\Delta\omega)_i(T)_{qh}/\omega_i(0) = -\gamma_i(\Delta V(T)/V(0))$. The Grüneisen parameter calculated for RMnO$_3$ [R = Sm, Nd and Pr] is ~ 2 [30, 31]. The quasi harmonic contribution in TbMnO$_3$ can be neglected since the fractional change in volume is negligible [32], as has been done in earlier studies of rare earth manganites RMnO$_3$ (R = Gd, Eu, Pr, Nd, Sm, Tb, Dy, Ho, and Y) [20, 22].

In a cubic anharmonic process, a phonon of frequency $\omega(\vec{\kappa}=0)$ decays into two phonons $\omega_1(\vec{\kappa_1})$ and $\omega_2(\vec{\kappa_2})$, keeping energy and momentum conserved, i.e. $\omega = \omega_1 + \omega_2$, $\vec{\kappa_1} + \vec{\kappa_2} = 0$. Considering the simplest decay channel with $\omega_1 = \omega_2$, the temperature dependence of ω (T) and the full width at half maximum (Γ) can be expressed as [33]

$$\omega(T) = \omega(0) + C\,[1+2n(\omega(0)/2)] \qquad (2)$$

$$\Gamma(T) = \Gamma(0) + A\,[1+2n(\omega(0)/2)] \qquad (3)$$

where n($\omega$) = 1/(exp($\hbar\omega/\kappa_B T$) -1) is the Bose–Einstein mean occupation number and C and A are the self-energy parameters for a given phonon mode. We realize that below the phase transition temperature T$_N$, eqs (2) and (3) are not expected to hold good, as is obvious in the temperature dependence of S5 mode (see Fig. 2a). Therefore, we fit the data between 50K to 300K using eqs (2) and (3) and the theoretical curves are extrapolated below 50K using the fitted parameters given in Table I (see solid lines in Fig. 2). Similar procedure has been adopted in earlier studies of manganites [20, 22]. We do not observe any significant signature, within the accuracy of our experiments, of the ferroelectric transition at T$_{FE}$ (~ 27K) in the temperature dependence of frequencies and linewidths. The fit of the data with Klemens model is modest. The fitting parameter C of mode S6 is very high as compared to the other modes, showing that this mode is much more anharmonic. The linewidths and frequencies of the modes S3 and S4 show normal temperature dependence and are not shown here.

An interesting observation is that the intense mode S5 shows anomalous temperature dependence: the mode shows softening near T$_N$ ~ 46K. Similar anomalous temperature dependence has been observed for a few Raman modes in RMnO$_3$ where R = La [21] and Gd, Pr, Nd, Sm, Dy [20, 22], which has been attributed to spin-phonon coupling [21]. This is understood as follows: if an ion is displaced from its equilibrium position by "*u*", then the crystal potential is given as U = 0.5* (k$u^2$) +$\Sigma_{ij}$ $J_{ij}(u)S_iS_j$, where k in the first term represents the force constant and the second term arises from spin interactions between the $Mn^{3+}$ spins. The phonon frequency is affected by the additional term $(\Delta\omega)_{sp-ph}$(T) = $\lambda$ <$S_iS_j$>, where $\lambda$ = ($\partial^2 J_{ij}(u)/\partial u^2$) is the spin-phonon coupling coefficient and <$S_iS_j$> is the spin-correlation function. The parameter $\lambda$ can be positive or negative and can be different for different phonons. Below $T_N$, the spin correlations build up and hence the spin-phonon coupling becomes important at lower temperatures. The renormalization of the mode S5 frequency starts slightly above $T_N$ (~ 46K), which can arise from spin fluctuations due to quantum and thermal effects [20].

**(B) Orbiton-Phonon Coupling**

We now discuss the two high energy excitations, S7 at (1156 $cm^{-1}$) and S8 at (1328 $cm^{-1}$). Figure 3 shows temperature-dependence of the frequency and linewidth of S8 mode as well as the intensity ratio of S8 and S6 modes (S8/S6) in the temperature range from 5K to 200K. Above 200K, the mode S8 is too weak to be analyzed quantitatively as is the case for S7. Mode S7 can be assigned as the second order Raman mode involving a combination of S2 and S6 or S4 and S5 phonons and S8 as a overtone of S6 (658 $cm^{-1}$) mode. As the second order Raman scattering involves the phonons over the entire Brillouin zone, the frequencies of the observed second order phonons are not necessarily the double of first order phonons at the $\Gamma$- point ( q = 0,0,0). The intensity ratio of S8 to S6 is most interesting, namely, it is ~ 0.6 at all temperatures. The intensity ratio of S7 to S5 is ~ 0.1 and S7 to S6 is ~ 0.4 in the temperature range of 5K to 140K. This anomalously large intensity ratio even at low temperatures can only be understood by invoking the mixing of the multiphonon modes with the orbitons [1-2, 34]. Figure 3 also shows temperature dependence of the frequency and the linewidth of the S8 mode which has yet to be understood quantitatively for the mixed multiphonon-orbiton mode. It will be interesting to explore the role of spin-charge-lattice coupling in understanding multiphonon Raman scattering in multiferroic $TbMnO_3$. This suggestion arise from the observation of resonance Raman scattering from two-magnons in multiferroic $BiFeO_3$ wherein the resonance involves electronic levels, magnons and phonon states [35]. We hope our results will motivate further theoretical studies on this aspect.

## 4. Conclusions

In summary, we have carried out a detailed temperature dependence of the first and second order Raman modes in TbMnO$_3$. The intensity ratio of the second-order phonon (S8) to its first-order counterpart (S6) is unusually high and it remains constant down to 5K. This anomalous temperature dependence of the intensity ratio is attributed to the mixing of the second-order phonon with the orbitons as theoretically predicted. Four first-order modes (S1, S2, S3 and S4) show normal behavior with temperature, whereas the S5 mode behaves anomalously below T$_N$ probably arising from a strong spin-phonon coupling. We submit that the present study brings out yet another example of orbiton mediated multiphonon Raman scattering in the manganite family.

**Acknowledgments**

AKS thanks the Department of Science and Technology, India, for financial support. PK thanks Council of Scientific and Industrial Research, India, for research fellowship.

## References:

[1] P. B. Allen and V. Perebeinos 1999 Phys. Rev. Lett. **83**, 4828.

[2] V. Perebeinos and P. B. Allen 2001 Phys. Rev. B **64**, 085118.

[3] E. Saito, S. Okamoto, K. T. Takahashi, K. Kobe, K. Yamamoto, T. Kimura, S. Ishihara, S. Maekawa and Y. Tokura 2001 Nature **410**, 180.

[4] J. Vanden Brink 2001 Phys. Rev. Lett. **87**, 217202.

[5] R. Kruger, B. Schulz, S. Naler, R. Rauer, D. Budelmam, J. Backstrom, K. H. Kim, S. W. Cheong, V. Perebeinos and M. Rubhausen 2004 Phys. Rev. Lett. **92**, 097203.

[6] L. Martin-Carron and A. de Andres 2004 Phys. Rev. Lett. **92**, 175501.

[7] K. Y. Choi, P. Lemmens, T. Sahaouri, G. Guntherodt, Y. G. Pashkevich, V. P. Gnezdilov, P. Reutler, L. Pinsard-Gaudart, B. Bucher and A. Revcolevschi 2005 Phys. Rev. Lett. **71**, 174402.

[8] K. Y. Choi, P. Lemmens, G. Guntherodt, Y. G. Pashkevich, V. P. Gnezdilov, P. Reutler, L. P. Gaudart, B. Bucher and A. Revcolevschi 2005 Phys. Rev. B **72**, 024301.

[9] P. B. Allen and V. Perebeinos 2001 Nature **410**, 155.

[10] M. Gruninger, R. Ruckamp, M. Windt, P. Reutler, C. Zobel, T. Lorenz, A. Freimuth and A. Revcolevschi 2002 Nature **418**, 39.

[11] J. Bala, A. M. Oles and G. A. Sawatzky 2002 Phys. Rev. B **65**, 184414.

[12] M. N. Iliev, V. G. Hadjiev, A. P. Litvinchuk, F. Yen, Y. Q. Wang, Y. Y. Sun, S. Jandl, J. Laverdiere, V. N. Popov and M. M. Gospodinov 2007 Phys. Rev. B **75**, 064303.

[13] T. Kimura, T. Goto, H. Shintani, K. Ishizaka, T. Arima and Y. Tokura 2003 Nature **426,** 55.

[14] M. Bastjan, S. G. Singer, G. Neuber, S. Eller, N. Aliouane, D. N. Argyrion, S. L. Cooper and M. Rubhausen 2008 Phys. Rev. B **77,** 193105.

[15] A. Pimenov, A. A. Mukhin, V. Y. Ivanov, V. D. Travkin, A. M. Balbashov and A. Loidl 2006 Nature Physics **2**, 97.

[16] R. V. Aguilar, A. B. Sushkov, C. L. Zhang, Y. J. Choi, S. W. Cheong and H. D. Drew 2007 Phys. Rev. B **76**, 060404.

[17] Y. Takahashi, N. Kida, Y. Yamasaki, J. Fjioka, T. Arima, R. Shimano, S. Miyahara, M. Mochizuki, N. Furukawa and Y. Tokura 2008 Phys. Rev. Lett. **101**, 187201.

[18] R. V. Aguilar, M. Mostovoy, A. B. Sushkov, C. L. Zhang, Y. J. Choi, S. W. Cheong and H. D. Drew 2009 Phys. Rev. Lett. **102**, 047203.

[19] A. Pimenov, A. Shuvaev, A. Loidl, F. Schrettle, A. A. Mukhin, V. D. Travkin, V. Yu Ivanov and A. M. Balbashov 2009 Phys. Rev. Lett. **102**, 107203.

[20] J. Laverdiere, S. Jandl, A. A. Mukhin, V. Yu. Iyanov, V. G. Iyanov and M. N. Iliev 2006 Phys. Rev. B **73**, 214301.

[21] E. Granado, A. Garcia, J. A. Sanjurjo, C. Rettori, I. Torriani, F. Prado, R. D. Sanchez, A. Caneiro and S. B. Oseroff 1999 Phys. Rev. B **60**, 11879.

[22] W. S. Ferreira, J. A. Moreira, A. Almeida, M. R. Chaves, J. P. Araujo, J. B. Oliveira, J. M. Machado Da Silva, T. M. Sa, T. M. Mendonca, P. S. Carvalho, J. Kreisel, J. L Riberio, L. G. Vieira, P. B. Tavares and S. Mendonca 2009 Phys. Rev. B **79**, 054303.

[23] P. Kumar, S. Saha, C. R. Serrao, A. K. Sood and C. N. R. Rao 2010 Pramana J. Phys. (arXiv:0905.3092).

[24] L. Martin-Carron, A. de Andres, M. J. Martinez-Lope, M. T. Casais and J. A. Alonso 2002 Phys. Rev. B **66**, 174303.

[25] M. N. Iliev, M. V. Abrashev, J. Laverdiere, S. Jandl, M. M. Gospodinov, Y. Q. Wang and Y. Y. Sun 2006 Phys. Rev. B **73**, 064302.

[26] L. Martin-Carron, J. S. Benitz and A. de Andres 2003 J. Solid State Che. **171**, 313.

[27] V. B. Podobedov, A .Weber, D. B. Romero, J. P. Rice and H. D. Drew 1998 Phys. Rev. B **58**, 43.

[28] M. V. Abrashev, A. P. Litvinchuk, M. N. Iliev, R. L. Meng, V. N. Popov, V. G. Ivanov, R. A. Chakalov and C. Thomsen 1999 Phys. Rev. B **59**, 4146.

[29] M. N. Iliev, M. V. Abrashev, V. N. Popov and V. G. Hadjiev 2003 Phys. Rev. B **67**, 212301.

[30] R. Choithrani, N. K. Gaur and R. K. Singh 2009 J. Alloys and Compounds. **480**, 727.

[31] R. Choithrani, N. K. Gaur and R. K. Singh 2008 J. Phys.: Condens. Matter **20**, 415201.

[32] D. Meier, N. Aliouane , D. N. Argyriou , J. A. Mydosh  and T. Lorenz 2007 New J. Phys. **9**, 100.

[33] P. G. Klemens 1966 Phys. Rev. **148**, 845.

[34] Light Scattering in solids II, Topics in Applied Physics vol. 50, edited by M. Cardona (Springer-Verleg, Berlin, 1982).

[35] M. O. Ramirez, A. Kumar, S. A. Denev, Y. H. Chu, J. Siedel, L. W. Martin, S.-Y. Yang, R. C. Rai, X. S. Xue, J. F. Ihlefeld, N. J. Podraza, E. Saiz, S. Lee, J. Klug, S. W. Cheong, M. J. Bedzyk, O. Auciello, D. G. Schlom, J. Orenstien, R. Ramesh, J. L. Musfeldt, A. P. Litvinchuk and V. Gopalan 2009 App. Phys.Lett. **94**, 161905.

Table1: List of the experimental observed phonons frequencies and fitting parameters of a few phonons, fitted by equation (2) and (3) as described in text. Units are in cm$^{-1}$.

| Assigment | Phonon ω (5K) | Frequency ω(0) | C | Γ(0) | A |
|---|---|---|---|---|---|
| S1 ($A_g$) | 384.9 | 386.1 ± 0.6 | -4.1 ± 0.8 | 10.8 ± 2.7 | 5.9 ± 2.8 |
| S2 ($A_g$) | 491.9 | 495.3 ± 0.3 | -8.5 ± 0.5 | 12.3 ± 0.9 | 10.4 ± 1.4 |
| S3 ($A_g$) | 514.5 | | | | |
| S4 ($B_{2g}$) | 531.3 | | | | |
| S5 ($B_{2g}$) | 616.2 | 620.8 ± 0.7 | -6.2 ± 1.2 | 10.9 ± 1.2 | 14.1 ± 2.1 |
| S6 | 658.4 | 668.8 ± 4.9 | -27.2 ± 8.4 | 40.6 ± 3.2 | 0.09 |
| S7 (Second order) | 1157.1 | | | | |
| S8 (Overtone) | 1327.4 | | | | |

**Figure Captions:**

Figure1: (Colour online) Unpolarised-Raman spectra of TbMnO$_3$ measured at 5 K. Thick solid line shows the total fit, thin solid line show the individual Lorentzian fit.

Figure 2: (Colour online) Temperature dependence of the modes S1, S2, S5 and S6. The solid lines are the fitted curve as described in text.

Figure 3: Temperature dependence of the intensity ratio of the mode S8 to S6, S7 to S6 and S7 to S5 (Top Panel). Temperature dependence of the Frequency (Lower Panel) and Linewidth (Middle panel) of mode S8. Solid lines are guide to the eye.

**Figure1**:

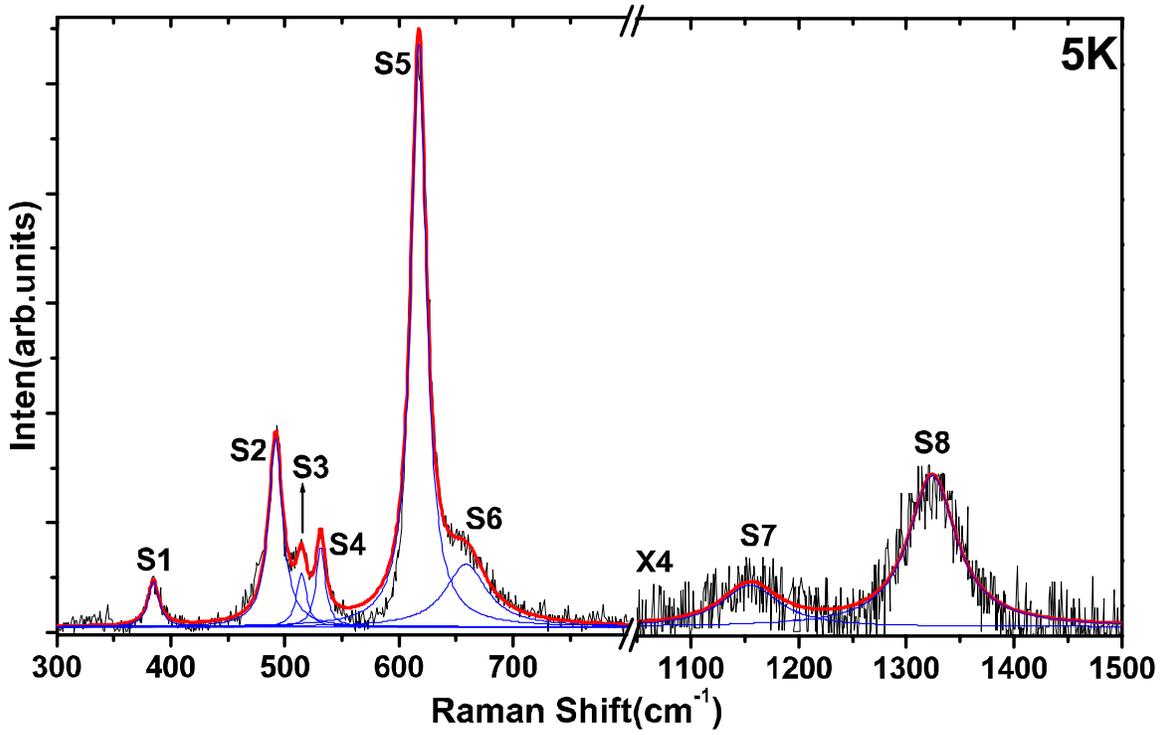

**Figure 2:**

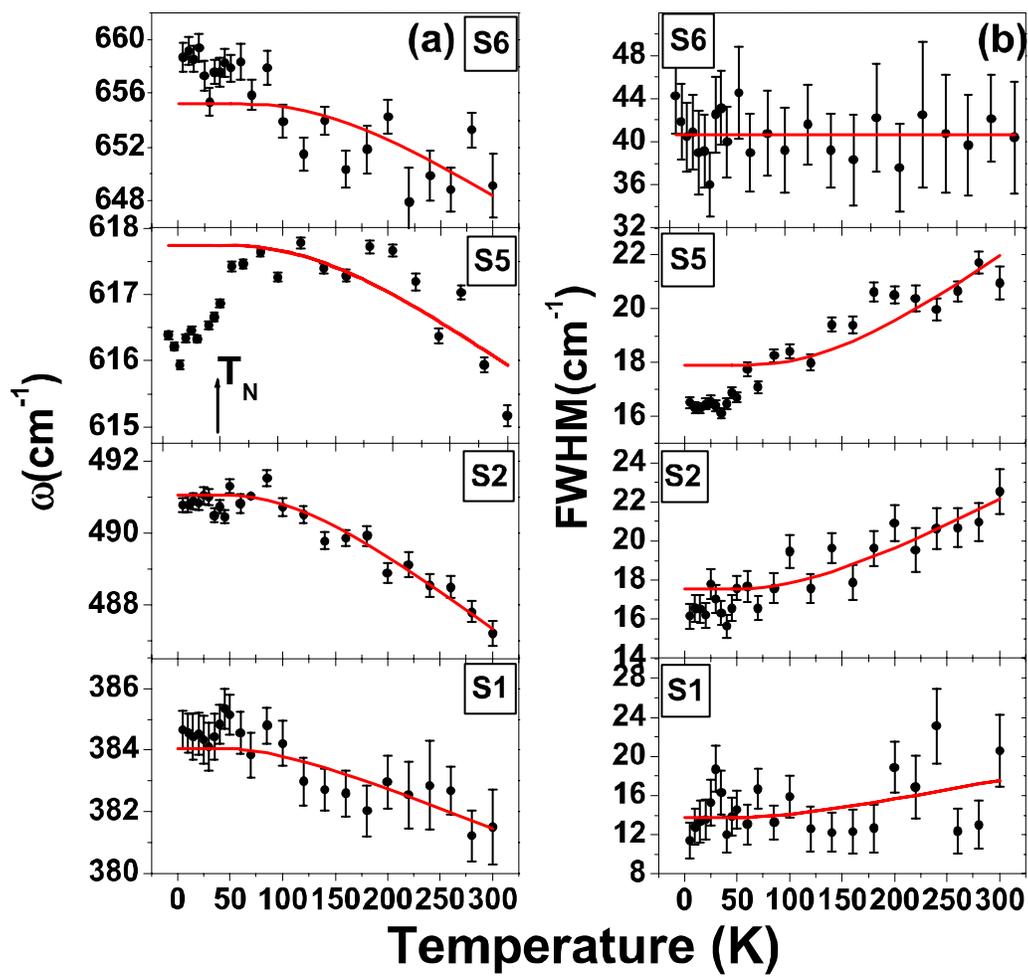

**Figure 3:**

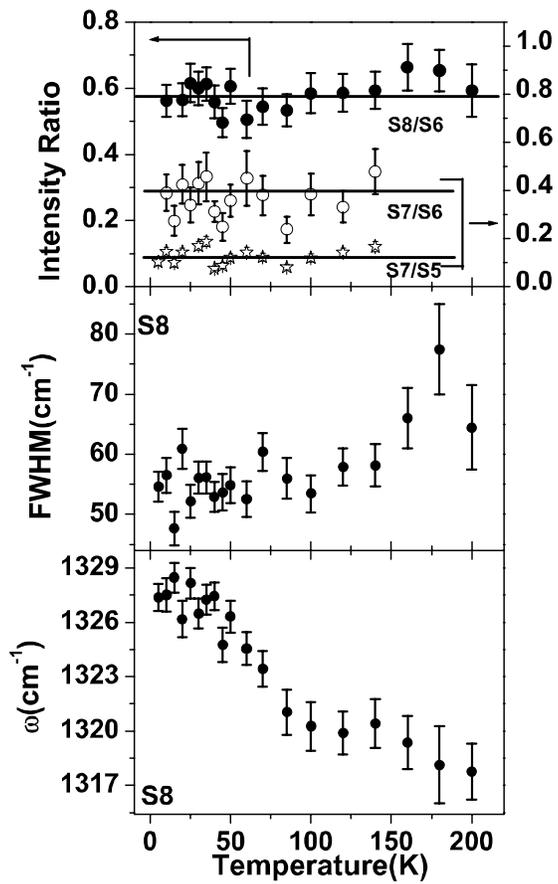